\documentclass[aps,prb,amsmath,amssymb,
showpacs,eqsecnum,twocolumn,floatfix]{revtex4}
\usepackage{graphicx}% Include figure files
\usepackage{dcolumn}% Align table columns on decimal point
\usepackage{bm}% bold math

\def\lsim{\mathrel{\rlap{\lower4pt\hbox{\hskip1pt$\sim$}}
    \raise1pt\hbox{$<$}}}                % less than or approx. symbol
\def\gsim{\mathrel{\rlap{\lower4pt\hbox{\hskip1pt$\sim$}}
    \raise1pt\hbox{$>$}}}                % greater than or approx. symbol

\def\jetp{{\em Sov. Phys. JETP~}}

\def\fig{Fig.~}
\def\eq{Eq.~}
\def\eqs{Eqs.~}

\begin{document}

\title{Tunneling conductance in Superconductor/Ferromagnet junctions:
a self consistent approach.}

\author{Paul H. Barsic}
%\affiliation{School of Physics and Astronomy, University of Minnesota,
%Minneapolis, Minnesota 55455}
 %\altaffiliation[Also at ]{Minnesota Supercomputer Institute, University of
 %Minnesota, Minneapolis, Minnesota 55455}
 \email{pbarsic@arete.com}
 \altaffiliation[Present address: ] {Aret\'{e} Associates, 1550 Crystal 
Dr. Ste. 703, Arlington,
Virginia 22202} 

\author{Oriol T. Valls}
\email{otvalls@umn.edu}
\affiliation{ School of Physics and Astronomy, University of Minnesota,
Minneapolis, Minnesota 55455}
 \altaffiliation[Also at ]{Minnesota Supercomputer Institute, University of
Minnesota, Minneapolis, Minnesota 55455}

\date{\today}

\begin{abstract}%otv1 abstract edited
We evaluate the tunneling conductance of clean Ferromomagnet/Superconductor
junctions via  a fully self-consistent numerical solution of the
microscopic Bogoliubov-DeGennes equations. We present results for
a relevant range of values of the Fermi wavevector mismatch (FWM),
the spin polarization, and the interfacial scattering strength. 
For nonzero spin polarization, 
the conductance curves
vary nonmonotonically with  FWM.
The FWM dependence of the self-consistent
results  is stronger than that previously found in non-self-consistent
calculations,
since, in the self-consistent case, the effective scattering potential near
the interface depends on the FWM. The dependence on interfacial
scattering is monotonic. 
These results confirm that it is impossible to characterize
both the the FWM and the interfacial scattering by a single 
effective parameter
and that analysis of experimental data via the use of  such 
one-parameter models
is  unreliable.

\end{abstract}

\maketitle

\section{Introduction}

Refinements
in fabrication techniques that have occurred
over the past twelve years
have made it possible
to create devices  that exploit and display
the
interplay between ferromagnetic and superconducting orderings in 
superconductor/ferromagnet (SF) heterostructures. 
These developments have raised the possibility %otvr
of building devices that may be used to manipulate and detect
spin polarized currents. Such devices, besides being %otvr
of obvious considerable scientific
interest, have technological applications for 
spintronics.\cite{zutic04} 
It has been suggested that the introduction of an F layer into
solid-state qubits may allow a stabilization of the state of the junction
without the need for an external field.\cite{blatt}  
Study of SF bilayer devices may illuminate the %phb3
behavior of other systems that undergo spin/charge separation.\cite{kashi}
Spin/charge separation occurs when an electron
is injected into a superconductor at the gap edge. The charge is absorbed by 
the condensate and quickly carried away by the supercurrent while its 
spin excitation remains.\cite{leridon}  
%In SF junctions, this leads to spin/charge separation, while the %phb
The different spin orderings of the S and F layers lead to spin dependent
transport effects, %. %% that may be used to manipulate spin polarized currents,
%making them good candidates for spintronics applications, such as spin valves
%or spin polarization detectors.\cite{zutic04}  
providing ways to study phenomena %phb3
%Studies of SF junction conductivity provide ways of
%studying phenomena
such as  spin polarization, spin-diffusion lengths, and
spin-relaxation times.\cite{yama02,soulen} %%  leading to a better understanding of
Other potential applications and device geometries are discussed
elsewhere.\cite{zutic04, buzdin05, hv05}  Earlier work is extensively reviewed
in Ref.~\onlinecite{zutic00}. The knowledge gained from SF spin probes can lead, if
the results are properly analyzed, to a better understanding of %otv3 rewritten and moved
spin transport properties in different materials and nanostructures.

Thanks to the new technologies mentioned above, %phb
transport in clean SF bilayers is a very active subject of experimental 
research, but it remains a difficult topic theoretically.  
The problem of calculating the conductance of an S/F bilayer
has been solved,\cite{zutic00,zutic99} but only in a non-self-consistent 
manner, that is, by assuming the superconducting order parameter to
behave as a sharp step function at the interface. This assumption neglects the
proximity effect, the ``leaking'' of superconductivity into the ferromagnet %otv3
(and, {\em vice versa}, that of the magnetism into the superconductor) %otv3
that occurs in real systems. Thus, the proximity effect makes %phb3
it impossible to identify a
precise  location where the superconducting
correlations end and
the ferromagnetic ones begin. %otv3 moved
Even with this simplification, %still leaves %phb
the non-self-consistent approach is still far from %phb
trivial, as one must 
consider Andreev reflection, \cite{andreev}
the spin asymmetries due to the exchange interaction, 
and the spin coupling at the interface. 
Nevertheless, the stark fact remains that neglecting %otvr %phb
a self-consistent order parameter
amounts to no less
than omitting all influence of S/F  proximity effects.
We remedy this situation in this paper, where we show how the
difficulties associated with the proximity
effect can be overcome and how a self-consistent calculation %otvr
of the conductance can be achieved.  
The inclusion of the order
parameter in a fully  self-consistent manner
allows us to fully account for the influence of the %otvr
proximity effect, leading to  correct, accurate and %otvr
more interesting results at the worthwhile cost of having to solve a %otvr
much more complicated problem.  
The self-consistent procedure results in the  physical situation of a %otvr
scattering potential that is dependent on the wavefunctions of the scattered
particles and holes.  There is no analytic approach that will give a fully
self-consistent pair amplitude for this problem: %otvr
%gives us a numerical rather than an analytic form of the scattering potential, 
a numerical approach must be taken. %otvr
%The
%Andreev scattering at the interface is an effect of the superconducting 
%gap
%density of states (DOS).
The spatial variations in the pair amplitude complicate the details of Andreev
reflection\cite{andreev} 
%by effectively introducing a spatially varying scattering potential.
since the density
of states (DOS) can no longer be described as having a well defined BCS 
gap,\cite{hv04a}  % is this the one where they talk about DOS?  See about it...
and one must consider sub-gap states and even gapless superconductivity.
Since there is no exact analytic solution for the spatial variations
in the pair amplitude at an SF interface, there is no analytic solution 
for the DOS either, nor {\it a fortiori} for the conductance. %otv3
In short, this is a complicated four-component
scattering problem without an analytic form for the scattering potential.
Despite these difficulties, the proximity effect 
must not be neglected, as it has, as we shall see,  a strong influence on 
the results and on the way experimental
data must be analyzed. %otvr

Early experimental work\cite{mess} was done on devices with an insulating oxide layer
between the S and F layers, thus incorporating %otvr
a superconductor/insulator/ferromagnet %phb
(SIF) tunnel junction.  
The stable oxide layer prevented the diffusion of ferromagnetic impurities into
the superconductor, making it possible to work in the clean limit.  
One disadvantage of SIF systems is that a
%Since the long range decay proximity effect is found in SN systems and not I/S
%systems, a thin 
tunnel barrier inhibits the proximity effect.
%for tunneling and allowing the
%possibility of a coexistence of ferromagnetic and superconducting orders.
Another disadvantage of the SIF junction is the reduced spin
coherence lengths.  While the electrons merely tunnel through the insulator, 
the strong binding to the lattice leads to spin decoherence.  These  %phb
tunnel junctions provide good information on the superconducting DOS, but it is
dubious whether they provide a good picture of the state of the ferromagnet near %otvr
the interface, and they will certainly not reveal the consequences of
the proximity effect.
Other work\cite{johnson} in a spin transistor geometry demonstrated
the possibility of growing SF interfaces over a large area 
with no insulating oxide layer.  The purpose of that study was to
explore transverse spin currents, not to examine bilayer conductance at the SF 
interface.
Also, it is not clear that there was no diffusion of magnetic
atoms at the interface.  
More recent\cite{reymond,hacohen} work has focused on bilayer conductance in
%has shown the possibility of growing large 
planar junctions with sharp interfaces and clean materials.
Characteristic peaks in the conductance were observed at bias voltages
corresponding to $\Delta_0$, the bulk superconducting gap energy, and an 
enhanced conductance at zero bias.

Point contact bilayers offer a convenient method for studying %otv3
relatively abrupt 
interfaces.\cite{raych,shashi,perez,chalsani}  
These devices are made by growing an oxide layer, thick enough to
suppress tunneling, on top of a planar ferromagnet.
A small hole is 
made through the oxide, and a superconductor is grown on top.  
This geometry prevents
diffusion of ferromagnetic atoms during growth and allows for a uniform
magnetic field.  In this way, it is possible to study
experimentally\cite{raych,shashi} the conductance of
clean SF bilayers in the
ballistic limit.
A quicker method for making point contacts mechanically places the tip of a sharpened
superconducting wire on a bulk F sample.\cite{soulen}  While such
a technique
gives poor control over the size and shape of the contact, 
it is technologically a very desirable procedure as it leads
to quickly obtained results, and therefore
it may be used to 
probe the spin states of many different types of F materials.
However, the analysis  of the data,
and of conductance spectroscopy experimental results generally, has
been hampered by incomplete  understanding of how to
relate data and theory. 
We will extensively address this issue in this work. 
We will be able to show that there is a clear
influence of the proximity effect on transport at SF interfaces, which
will enhance our understanding of these devices.  This will also show that
fully self-consistent studies are necessary to actually understand these point
contact devices.

In the case of strong interfacial scattering, Andreev reflection and the
proximity effect are suppressed.  In the
limit of a very large interfacial barrier, 
the conductance reflects essentially the superconducting %otvr
DOS.  If we work with small or medium
interfacial barriers, as is the case here, 
Andreev reflection and the proximity effect become very important,
leading to a much more interesting but also much more difficult problem.  
In the small barrier limit Andreev reflection
can lead to an enhancement of sub-gap conductance\cite{zutic00}
since a single electron from F must excite a pair of electrons in S.  This
agrees with experimental observations of zero bias
conductance.\cite{raych,hacohen,shashi,reymond}  The proximity effect changes the local
DOS\cite{hv03} in the vicinity of the junction, introducing sub-gap states and
even gapless superconductivity. %, taking us further away from the SIF result.

Semiclassical methods,
such as the Eilenberger or Usadel equations,\cite{buzdin05,kraw02,leadbeater,seviour}% or some
%application of the Boltzmann transport equation.\cite{yama02}
can give a reasonable approximation to the conductance 
curves\cite{pepe} in dirty systems. However, such models are not 
appropriate for clean
systems, and can lead to spurious predictions.\cite{reymond,pepe}  A common
phenomenological approach is to define a current polarization
parameter\cite{shashi} based upon the difference in spin-up and spin-down DOS
in F.
The portion of the transmission coefficient due to the Andreev reflected hole
(AR hole) is calculated for an SN interface.  Since the AR hole must be in the
opposite spin band of the incident electron, the AR hole coefficient is
modified by a simple function of the polarization parameter.
%This is a not obviously unreasonable method of analysis,% phb
This is a reasonable phenomenological approach, %phb
but it lacks any microscopic
justification, and is usually of limited success.\cite{perez}

There have been many attempts to derive a fully microscopic model to describe
the conductance in SF 
systems.\cite{kashi,zutic00,zutic99,shashi,dejong,stefanakis,melin,tkachov,cayssol,bozovic}  
These studies have used an abrupt approximation for the pair
amplitude at the interface, thereby focusing on elastic scattering and Andreev
reflection but neglecting the proximity effect.
These studies do predict some correct qualitative features of the junctions. 
The author of Ref.~\onlinecite{stefanakis} remarks, and we agree, that
only a proper consideration of the proximity effect through self-consistent
methods will give the correct quantitative features as well.
A self-consistent study\cite{kuboki} 
relying on a 2-D tight binding model has been performed but only to 
study spatial variations of the spin and charge currents parallel to interface.
Therefore, there is a need for a fully self-consistent microscopic 
treatment of the conductance in SF bilayers to obtain quantitatively 
accurate results. %otvr
%treatment for the conductance of SF bilayers.

In this work we show how to  numerically calculate conductances of 
SF bilayers using a
fully self-consistent solution to the Bogoliubov deGennes (BdG) equations
with  a net current. %otv3
We require a fully self-consistent solution to
the BdG equations to properly treat clean 
inhomogeneous systems in three dimensions.\cite{zutic00} 
%We will analyze the resulting eigenfunctions, with proper boundary conditions, %otv3
%and generate from them  
%via the Blonder-Tinkham-Klapwijk (BTK) method,\cite{btk82} %otv3
%which provides the best approach to extracting the %otv3
%conductances from a fully microscopic model
%describing scattering in
%the ballistic limit,
%transmission probabilities for particles scattered from F to S. %otv3
We will analyze the resulting eigenfunctions, with proper boundary conditions, 
and generate transmission probabilities for particles scattered from F to S
via the Blonder-Tinkham-Klapwijk (BTK) method.\cite{btk82} %otv3 %phb4
%The BTK method provides the best approach to extracting the conductances from a 
%fully microscopic model describing scattering in the ballistic limit.  %phb4
These transmission probabilities can then be used to calculate I-V curves, %otv3
from which we
will calculate the conductances.
We find good qualitative agreement with
experimental results, including an enhanced conductance at zero bias. 
We show that  the Fermi wavevector mismatch at the interface, the
exchange field in the ferromagnet, and the interface barrier
scattering all have a 
significant and
independent effect on the shape of the conductance curves.
This paper represents the early fruits of a completely new technique, which
we expect
will eventually be used to study more complicated
geometries and conditions. 
Our fully self-consistent solutions will allow for a very careful consideration
of the influence of the proximity effect on the conductance, even for very
thick F layers.

In the next section, we describe in some detail our numerical methods %phb3
and the procedures that we follow to extract the conductance as a function
of applied voltage. The results are presented and discussed in detail in 
Sect.~III. A brief final section recapitulates  our conclusions and points
to future directions.

\section{Methods}

The systems we study here are planar junctions made of SF bilayers with
atomically smooth interfaces.  We assume
that the layers are
semi-infinite in the directions perpendicular to the interfaces
(the $x-y$ directions). 
We will take the bands to be parabolic, thus %otv3 
$\epsilon_\perp={k_\perp^2}/{2m}$, where $k_\perp$ is the wavevector in the
transverse direction and $\epsilon_\perp$ is the energy corresponding to the
$x-y$ variables. The superconductor is assumed to be an $s-$wave material.

We use a numerical
diagonalization of
the microscopic Bogoliubov-deGennes\cite{dg}  equations
for this inhomogeneous system.  Given a pair potential (order parameter)
$\Delta(z)$, to be determined self
consistently, the spin-up quasi-particle ($u_n^\uparrow(z)$) and spin-down 
quasi-hole ($v_n^\downarrow(z)$) amplitudes obey the BdG
equations:
\begin{widetext}
\begin{equation}
\left( \begin{array}{cc}
H-h(z) & \Delta(z) \\
\Delta(z) & -(H+h(z)) \end{array} \right)
 \left( \begin{array}{c}
u_n^\uparrow(z) \\
v_n^\downarrow(z)
\end{array} \right)
= \epsilon_n
 \left( \begin{array}{c}
u_n^\uparrow(z) \\
v_n^\downarrow(z)
\end{array} \right).
\label{dgeq}
\end{equation}
\end{widetext}
Where $H={p_z^2}/{2m} - E_F(z) + \epsilon_\perp + U(z)$ is a  single-particle
Hamiltonian where ${p_z^2}/{2m}$ is the 
contribution to the kinetic energy from motion in the $z$ direction.
The  continuous variable $\epsilon_\perp$ 
is  decoupled 
from the $z$ direction but of course it
affects the eigenvalues $\epsilon_n$, 
which are measured from the chemical potential.
We describe the magnetism by an exchange field $h(z)$ which takes the value
$h_0$ in the F material and vanishes in S.
Within the superconducting layer, $E_F(z)$ is equal to $E_{FS}$, the Fermi 
energy of the S layers measured from the bottom of
the band, while in the ferromagnet we have
$E_F(z)=E_{FM}$ so that in the F regions the 
up and down band widths are $E_{F\uparrow}= E_{FM}+h_0$ and
$E_{F\downarrow}= E_{FM}-h_0$ respectively. 
%As pointed out  in Ref.~\onlinecite{zutic00} one should
As pointed out  in Ref.~\onlinecite{zutic00}, one should %phbr
not assume that $E_{FM}=E_{FS}$ and we therefore introduce the
dimensionless Fermi wavevector mismatch parameter % $\Lambda$ by %phbr
$\Lambda \equiv E_{FM}/E_{FS}\equiv (k_{FM}/k_{FS})^2$. 
We also introduce the dimensionless magnetic strength
variable $I$ by $h_0 \equiv E_{FM}I$. The $I=1$ limit corresponds
to the ``half-metallic'' case.  
Interfacial scattering is described by the potential $U(z)$ which
we take to be of the form $U(z)=H\delta(z-z_0)$ where $z_0$ is the
location of the interface. The dimensionless parameter $H_B\equiv mH/k_{FM}$
(everywhere in this paper $\hbar=1$) conveniently
characterizes the strength of the interfacial scattering.   
The amplitudes $u_n^\downarrow(z)$ and $v_n^\uparrow(z)$ 
can be written down from symmetry 
relations.\cite{dg} 
We will consider physically
relevant values of $\Lambda$,  $I$, and $H_B$. 
We measure all lengths in terms of the inverse of %otv3
$k_{FS}$, the Fermi wavevector in the S material.
The dimensionless F width in the $z$-dimension is $D_F=k_{FS}d_F$,
that of S is  $D_S=k_{FS}d_S$, and that of 
the entire sample is $D$.  The dimensionless spatial coordinates are denoted by
$Z=k_{FS}z$.
The pair potential $\Delta(Z)$  %otv3 no paragraph
must be found through the self-consistent condition:
\begin{equation}
\Delta(Z) = \frac{g(Z)}{2}
{\sum_n}^\prime
\left( u_n^\uparrow(Z)   v_n^\downarrow(Z)  +
       u_n^\downarrow(Z)   v_n^\uparrow(Z) \right)
\tanh\left(\frac{\epsilon_n}{2T}\right).
\label{self}
\end{equation}

We will show that we can treat this problem as a plane wave 
scattering problem, with a 
scattering potential that exists over a finite region of space.
The scattering potential is, of course, obtained through self-consistent 
methods.  We will then
look at the transmitted and reflected waves sufficiently far away from the
interaction region that they, too, are plane waves.\cite{merzbacher}  In this
case, the spatial extent of the scattering potential is governed by the
proximity effect.  Therefore, we must take the sample size large enough that
the pair amplitude is zero over a large fraction of F, and it is approximately
equal to its bulk value $\Delta_0$ 
over a large fraction of S.  We call these regions in which the pair amplitude
is approximately constant the asymptotic regions.
We must also take the total sample size to be very
large so that the minimum wavevector in the problem, $k_{min}=\frac{2\pi}{D}$,
allows us to approximate a continuum of incident plane waves.  In principle, we
should take the sample size to be infinite, but we must choose a finite 
%otv lines below limit
value for computational considerations.  %We choose $D_S=D_F=30\xi_0$.
%We take the coherence length, $\xi_0$, to be $50$ in units of $k_{FS}^{-1}$.
%As explained below,  %phb2
We have taken a total sample size of sixty
times the superconducting coherence length $\xi_0$, which turned out to
be sufficient to avoid finite size effects.

The procedure for calculating conductances begins by using the BdG equations
(\ref{dgeq}) and the self consistency condition
(\ref{self}) to
find a self-consistent order parameter for the system.  We use a procedure
similar to that in previous work.\cite{hv04a,bhv07,hv05,hv02a}
We start with a fully three dimensional
wavefunction $\Psi({\bf r})=e^{i{\bf k_\perp\cdot r}}\left(u(Z),v(Z)\right)^T$. The factor of 
$e^{i{\bf k_\perp\cdot r}}$ contributes only $\epsilon_\perp=k_\perp^2/2m$,
reducing this to a quasi-1D problem in the $Z$-direction. %otv3
We then expand the $u(Z)$ and $v(Z)$ eigenfunctions in a basis of both sines 
and cosines:
\begin{equation}
\phi_{q\pm}(Z)=\left\{
\begin{array}{ll}
\sqrt{\frac{2}{D}}\cos(k_q Z)& \\ %\mbox{if $0\le q \le N$}\\
\sqrt{\frac{2}{D}}\sin(k_q Z)& %\mbox{if $N< q \le 2N$}
\end{array}
\right.
\label{eq:basis}
\end{equation}  
where the $\pm$ signs in the left subindex refer to the sine or the cosine 
function respectively. 
This choice of basis is equivalent to using complex exponentials, but we gain some %otv3
computational advantage in the very time consuming step of calculating a
self-consistent pair amplitude by working with real rather than complex
numbers for the time being. The wavevectors $k_q$ are defined in %phb3
%We have for the wavevectors $q$ in
units of $k_{FS}$ as:
\begin{equation}
%k_q=\left\{
%\begin{array}{ll}
k_q=\frac{2 \pi q}{D} %& \mbox{if $0\le q \le N$}\\
%\frac{2 \pi (q-N)}{D}& \mbox{if $N< q \le 2N$}
%\end{array}
%\right..
\end{equation}  
with $q$ being a positive integer.
While the basis in principle requires all $q\ge 0$,
in practice it is sufficient to chose a  
a cut-off large enough so that the largest wavevector in the problem
corresponds to an energy that is a few $\omega_D$ above the Fermi level.
This choice of basis, which implies periodic boundary conditions as needed %otv3
in this problem, allows for a wavefunction that is non-zero %otv4 %phb4
and has a non-zero first derivative at the boundaries, two %phb3 
necessary conditions for a non-zero current to be present.   
%otv3 above said "this coice of basis" twice and "at the boundaties" thrice
\begin{widetext} %phb4 (necessary for change to 2 column)
In this basis, the upper left quadrant of the matrix in the left side of
Eqns.~(\ref{dgeq}) is:
\begin{eqnarray}
H_{q+p+}^+=\frac{2}{D}
(k_q^2+\epsilon_\perp+1)\delta_{pq}+&\frac{2}{D}(1-\Lambda-I)\left[
\frac{\sin((k_p-k_q)D_F)}{k_p-k_q}-\frac{\sin((k_p+k_q)D_F)}{k_p+k_q}
 \right] \nonumber \\
  & +\frac{4\sqrt{\Lambda}H_B}{D}(\cos(k_qD_F)\cos(k_pD_F)+1  ),
\end{eqnarray}
and the lower right quadrant is:
\begin{eqnarray}
H_{q+p+}^-=\frac{-2}{D}
(k_q^2+\epsilon_\perp+1)\delta_{pq}- & \frac{2}{D}(1-\Lambda+I)\left[
\frac{\sin((k_p-k_q)D_F)}{k_p-k_q}-\frac{\sin((k_p+k_q)D_F)}{k_p+k_q}
 \right] \nonumber \\ 
 & -\frac{4\sqrt{\Lambda}H_B}{D}(\cos(k_qD_F)\cos(k_pD_F)+1  )
\end{eqnarray}
for the cosine terms. %$p,q\le N$.
For the sine terms: %For  $p,q>N$,
\begin{eqnarray}
H_{q-p-}^+=\frac{2}{D}
(k_q^2+\epsilon_\perp+1)\delta_{pq}+ & \frac{2}{D}(1-\Lambda-I)\left[
\frac{\sin((k_p-k_q)D_F)}{k_p-k_q}-\frac{\sin((k_p+k_q)D_F)}{k_p+k_q}
 \right] \nonumber \\
  & -\frac{4\sqrt{\Lambda}H_B}{D}(\sin(k_qD_F)\sin(k_pD_F)  ),
\end{eqnarray}
and
\begin{eqnarray}
H_{q-p-}^-=\frac{-2}{D}
(k_q^2+\epsilon_\perp+1)\delta_{pq}- & \frac{2}{D}(1-\Lambda+I)\left[
\frac{\sin((k_p-k_q)D_F)}{k_p-k_q}-\frac{\sin((k_p+k_q)D_F)}{k_p+k_q}
 \right] \nonumber \\
 & +\frac{4\sqrt{\Lambda}H_B}{D}(\sin(k_qD_F)\sin(k_pD_F)  ).
\end{eqnarray}
For the cross terms involving a sine and a cosine we have: %$p>N$ and $q\le N$, we have:
\begin{eqnarray}
H_{q+p-}^+=\frac{2}{D} &
\left( \frac{\cos((k_p-k_q)D)-\cos((k_p-k_q)D_F)}{k_p-k_q}-
\frac{\cos((k_p+k_q)D)-\cos((k_p+k_q)D_F)}{k_p+k_q} \right)-\nonumber\\
&\frac{2}{D}(\Lambda-I) \left( \frac{\cos((k_p-k_q)D_F)-1}{k_p-k_q}-
\frac{\cos((k_p+k_q)D_F)-1}{k_p+k_q} \right) \nonumber \\
&-\frac{4\sqrt{\Lambda}H_B}{D}(\sin(k_qD_F)\cos(k_pD_F)  ).
\end{eqnarray}
and
\begin{eqnarray}
H_{q+p-}^-=-\frac{2}{D} &
\left( \frac{\cos((k_p-k_q)D)-\cos((k_p-k_q)D_F)}{k_p-k_q}-
\frac{\cos((k_p+k_q)D)-\cos((k_p+k_q)D_F)}{k_p+k_q} \right)-\nonumber\\
&-\frac{2}{D}(\Lambda+I) \left( \frac{\cos((k_p-k_q)D_F)-1}{k_p-k_q}-
\frac{\cos((k_p+k_q)D_F)-1}{k_p+k_q} \right)\nonumber\\
&+\frac{4\sqrt{\Lambda}H_B}{D}(\sin(k_qD_F)\cos(k_pD_F)  ).
\end{eqnarray}
In the above expressions, the matrix elements are given in 
units of $E_{FS}$, i.e. in dimensionless form. %otv fixed
\end{widetext} %phb4 (necessary for change to 2 column)

The self consistency condition, Eq.~(\ref{self}), requires us to %phb4
calculate the $\langle \phi_p(Z) | \Delta(Z) | \phi_q(Z)\rangle$ matrix elements
numerically. 
The iterative procedure is then
to cycle through Eqns.~(\ref{dgeq}) and (\ref{self}) %otv3
until convergence is obtained, as has been explained 
elsewhere.\cite{hv04a,bhv07,hv05,hv02a} The very large sizes required
for this problem, however, make the computations technically more
demanding: it is necessary to massively vectorize all programs in order
to obtain results in a reasonable amount of physical time.

\begin{figure}
%\centering{ \scalebox{1.0}{\includegraphics{btk.eps}  } 
\includegraphics [width=3.5in] {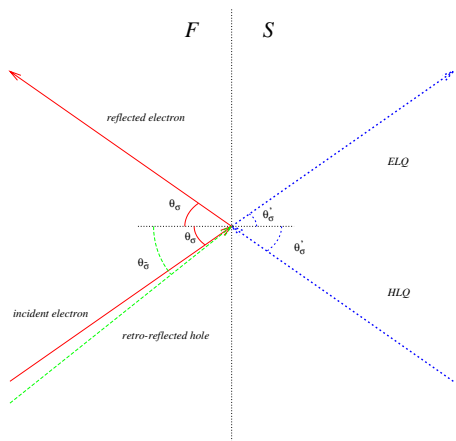}  
\caption{(Color online) Scattering at an SF interface.  An electron is incident from the
ferromagnet at an angle $\theta_\sigma$ with spin $\sigma$ and 
momentum ${\bf k_\sigma}$.  Ordinary scattering leads to a partially reflected
electron with the same spin and wavevector.  For sub-gap scattering,
excitations in the superconductor are Cooper pairs, consisting of an
electron-like quasiparticle (ELQ) and a hole-like quasiparticle (HLQ), carrying
a net spin of zero and charge $2e$.  To conserve charge, a hole is
retro-reflected (Andreev reflection) at an angle $\theta_{\bar{\sigma}}$
 into the opposite spin band, denoted by $\bar{\sigma}$.
}
\label{fig:btk}
\end{figure}

Once we have obtained a fully self-consistent spectrum of eigenfunctions and
eigenvalues,  we 
must extract a conductance from them.  
To successfully apply a plane wave scattering approach, we require a sample
large enough so that we can assume that the scattering potential is confined to
a finite region about the interface and that
there are ``asymptotic'' regions in S and F, far away from the scattering %otv1
region, where % it where %phbr %otv1
the scattering potential is not felt and the behavior is
bulk-like.  The
non-self-consistent work\cite{zutic00} assumes the scattering region to be infinitely %phbr
small.  
The general outlines for the self-consistent and non-self-consistent 
treatments are similar.
We will describe here first the salient features %otv3 changes here and below
of the non-self-consistent treatment, and then discuss the changes necessary 
to study the self-consistent case.

Consider a single electron of spin $\sigma$ in F with momentum ${\bf k_\sigma}$, %otv3
which is at an angle $\theta_\sigma$ with the $Z$-axis (see
\fig\ref{fig:btk}).  This electron may be partially reflected from the
interface as an electron with spin $\sigma$ and a
momentum of the same magnitude, but with the $k_z$ component in the opposite
direction (ordinary reflection).  
Andreev reflection allows a single charge from F to create a Cooper pair 
excitation in S without violating charge conservation.  In this process, 
when a single electron in spin band $\sigma$ is incident on the interface, 
a hole is retro-reflected into opposite spin band, which we denote by 
$\bar{\sigma}$, with momentum ${\bf k_{\bar{\sigma}}}$, and a Cooper pair
excitation is created in S.
%The transmitted particle is made of a pair of 
%electrons, with spin $\sigma$ and $\bar{\sigma}$.
%To allow for a transmission, $\sigma$ and an amplitude of unity 
%and no reflected particle of spin $\bar{\sigma}$.  
We write this as:
\begin{equation}
\psi_\sigma(Z)=
e^{i k_\sigma Z} \left( \begin{array}{c} 1\\0 \end{array} \right)
+a_\sigma e^{ik_{\bar{\sigma}}Z}\left( \begin{array}{c} 0\\1 \end{array}\right)
+b_\sigma e^{-i k_\sigma Z} \left( \begin{array}{c} 1\\0 \end{array} \right),
\label{psiF}
\end{equation}
where $k_\sigma$ and $k_{\bar{\sigma}}$ indicate the $Z$-components of 
${\bf k_\sigma}$ and ${\bf k_{\bar{\sigma}}}$, repectively.  As noted above
\eq(\ref{eq:basis}), %otv1
${\bf k_\perp}$ is conserved.  % %otv1 is this wrong? ${\bf k_\perp}$ is conserved. %phbr
We normalize to the incident particle flux, associate the ordinary reflected 
electron with
the $b_\sigma$ amplitude and the Andreev reflected hole with the $a_\sigma$
amplitude.  
The transmitted wavefunction in S is:
\begin{equation}
\psi^\prime_\sigma(z)=
c_\sigma e^{i k^\prime_\sigma z} \left( \begin{array}{c}
u_\sigma\\v_{\bar{\sigma}} \end{array} \right) +
d_\sigma e^{-i k^\prime_\sigma z} \left( \begin{array}{c}
v_{\bar{\sigma}}\\u_\sigma \end{array} \right)
\label{psiS}
\end{equation}
where $c_\sigma$ corresponds to an electron-like quasiparticle (ELQ) moving to 
the right and $d_\sigma$ corresponds to a hole-like quasiparticle (HLQ) moving 
to the left.  The $u_\sigma$ and $v_{\bar{\sigma}}$ amplitudes must obey the
normalization condition $u_\sigma^2+v_{\bar{\sigma}}^2=1$.
Similar equations can be written down for states with incident holes.
We can use the $a_\sigma$ and $b_\sigma$ amplitudes to write down a
formula for the conductance in the $T\rightarrow 0$ limit:\cite{zutic00}
\begin{eqnarray}
G(\varepsilon,\theta)&\equiv&
G_\uparrow(\varepsilon,\theta)+G_\downarrow(\varepsilon,\theta)\nonumber \\ &=&
\frac{e^2}{h}\sum_{\sigma} P_\sigma \left(
1+\frac{k_{\bar{\sigma}}}{k_\sigma}|a_{\bar{\sigma}}|^2-|b_\sigma|^2
\right),
\label{eq:cond}
\end{eqnarray}
where $P_\sigma=(1+\rho_\sigma I)/2$ accounts for the probability for the
incident electron to have spin $\sigma$,\cite{dejong}
%different DOS at the Fermi energy for the different spin bands,\cite{dejong}
 with $\rho_\sigma=1$ for the spin-up band and
$\rho_\sigma=-1$ for the spin-down band.
The ratio $\frac{k_{\bar{\sigma}}}{k_\sigma}$ accounts for the different spin
band wavevectors.
The conductance is a function of the incident angle of the electron
from F, $\theta\equiv\theta_\sigma$.  While we do not explicitly write it down, 
$a_{\sigma}$,  $b_\sigma$, $k_{\bar{\sigma}}$, and $k_\sigma$ are 
functions of $\varepsilon$ and $\theta$ as well.  
The angularly averaged conductance is given by\cite{zutic00}
\begin{equation}
G_\sigma(\varepsilon)\equiv\langle G_\sigma\rangle=\frac
{\int_0^{\Omega_\sigma}G_\sigma(\varepsilon,\theta)\cos(\theta)d\theta}
{\int_0^{\Omega_\sigma}\cos(\theta)d\theta},
\label{eq:aa}
\end{equation}
where $\Omega_\sigma$ is the angle of total reflection (critical angle) for 
incident particles of spin $\sigma$.

This model has the usual features of plane wave scattering.  The conservation
of $k_\perp$ across the interface leads to 
a modified\cite{zutic00} version of Snell's law:
\begin{equation}
k_{\sigma}\sin(\theta)=k_{\bar{\sigma}}\sin(\theta_{\bar{\sigma}})
\label{eq:snellR}
\end{equation}
and
\begin{equation}
k_{\sigma}\sin(\theta)=k_{\sigma}^\prime\sin(\theta_{\sigma}^\prime).
\label{eq:snellT}
\end{equation}
The incident and Andreev reflected angles are respectively given by
$\sin(\theta)=k_\perp/k_\sigma$ and 
$\sin(\theta_{\bar{\sigma}})=k_\perp/k_{\bar{\sigma}}$.  The transmitted angle
$\theta_\sigma^\prime$ is found through 
$\sin(\theta_\sigma^\prime)=k_\perp/k_\sigma^\prime$.
%, where $k_\sigma^\prime$ is the magnitude of the wavevector in S.  
This leads to a number of
phenomena similar to those in electromagnetic wave scattering. 
%but of note here is the critical angle phenomenon.  
In particular, the critical angle for ordinary reflection, which is given by 
$\Omega_\sigma=\sin^{-1}\left(\frac{k_\sigma^\prime}{k_\sigma} \right)
=\sin^{-1}\left(\frac{1}{\sqrt{\Lambda(1+\rho_\sigma I)}}\right)$,
 depends on the spin band of the incident electron.  There is also an angle
beyond which Andreev reflection is no longer possible, given by 
$\Omega_{A\sigma}=\sin^{-1}\sqrt{\left(\frac
{1+\rho_{\bar{\sigma}} I}
{1+\rho_\sigma I}
\right)}$.

The foregoing discussion applies also to the %otv3
non-self-consistent approach except that \eqs(\ref{psiF}) and (\ref{psiS}) %otv3
are now possible only in the asymptotic regions far from the interface. %otv3 
The comments that follow are the steps
necessary to analyze the self-consistent results.
Since there is no condition in the BdG equations to impose a net current
traveling to the right %otv3 in the asymptotic region, 
the self-consistent spectrum of eigenfunctions would not be, in the
asymptotic regions, in the convenient  %otv3
form of \eqs(\ref{psiF}) and (\ref{psiS}) even if we were to use a basis of
complex exponentials.  Our choice of periodic boundary conditions
allows us to impose the condition of a net current {\em a posteriori}.
This is very similar to the treatment given to one-dimensional scattering 
problems in elementary quantum mechanics.
Since a one-dimensional problem with periodic
boundary conditions will produce two-fold degenerate solutions,\cite{church}
%This is to our advantage because it will allow us to 
this quasi-one-dimensional problem also has two-fold degenerate solutions. 
We can find a linear combination of each pair of degenerate solutions  
that corresponds to a particle injected into the S layer
from the F layer.  If we look in the F layer sufficiently far away from the
interface (in the asymptotic region), so that the pair amplitude has gone 
completely to zero, we find that the numerically calculated
eigenfunctions can be fit to the form: %otv3 approximated by:
\begin{equation}  
\varphi_\sigma(Z)= \left( \begin{array}{c}
\eta_\sigma \sin(k_\sigma Z + \delta_\sigma)
\\ \eta_{\bar{\sigma}} \sin(k_{\bar{\sigma}} Z + \delta_{\bar{\sigma}})
\end{array} \right).
\label{eq:planewaveF}
\end{equation} 
%where the un-barred variables indicate one spin band, $\sigma$, and
%the barred variables indicate the other, $\bar{\sigma}$.
Similarly, in the asymptotic region of S, we find 
\begin{equation}  
\varphi_\sigma^\prime(Z)= \left( \begin{array}{c}
\eta_\sigma^\prime \sin(k_\sigma^\prime Z + \delta_\sigma^\prime)
\\ \eta_{\bar{\sigma}}^\prime \sin(k_\sigma^\prime Z +
\delta_\sigma^\prime)
\end{array} \right).
\label{eq:planewaveS}
\end{equation} 

If we take a degenerate pair of solutions, $\varphi_{\sigma A}(Z)$ and 
$\varphi_{\sigma B}(Z)$ , they will have the same
$k_\sigma$ and $k_{\bar{\sigma}}$ but different $\eta_\sigma$,
$\eta_{\bar{\sigma}}$, $\delta_\sigma$, and $\delta_{\bar{\sigma}}$ so
that they are orthogonal.  We can then find a unique linear
combination of a pair of degenerate states, 
$\psi_\sigma(Z)=A\varphi_{\sigma A}(Z) +
B\varphi_{\sigma B}(Z)$, that is in the form of \eqs(\ref{psiF}) and
(\ref{psiS}) in the asymptotic region.
The appropriate complex $A$ and $B$ coefficients are
\begin{equation}
A=\frac { \eta_{\bar{\sigma} B}e^{i\delta_{\bar{\sigma}A} } }
{
\eta_{\sigma A}\eta_{\bar{\sigma} B}e^{i(\delta_{\sigma A}+\delta_{\bar{\sigma}A})} - 
\eta_{\sigma B}\eta_{\bar{\sigma} A}e^{i(\delta_{\sigma B}+\delta_{\bar{\sigma}B})}  
},
\label{eq:LCA}
\end{equation}
and
\begin{equation}
      B=\frac{-A\eta_{\bar{\sigma} A}}
{\eta_{\bar{\sigma} B}e^{i(\delta_{\bar{\sigma}B}-\delta_{\bar{\sigma}A})}},
\label{eq:LCB}
\end{equation}
where we have subscripted the parameters to indicate the eigenfunction to which
they belong.  %In general, $A$ and $B$ are complex.
A small amount of elementary algebra  yields
expressions for the $a_\sigma$ and $b_\sigma$ coefficients, which we can then
substitute into \eq(\ref{eq:cond}) to find the conductance.  

To find quantities such as $k_\sigma$, $k_{\bar{\sigma}}$, $\eta_\sigma$,
$\eta_{\bar{\sigma}}$, $\delta_\sigma$, and $\delta_{\bar{\sigma}}$, we analyze
each eigenfunction in the asymptotic region of F.  The wavenumbers, 
$k_\sigma$ and $k_{\bar{\sigma}}$, are quantized in units of $k_{min}=2\pi/D$, 
by the discretization of the system.  To find their values in each case, 
we find the zeros of the eigenfunction to get the
characteristic wavenumber of the eigenfunction and assign it to the nearest
multiple of $k_{min}$.  
It is then a trivial step to obtain values for  $\delta_\sigma$, and 
$\delta_{\bar{\sigma}}$.  Finally, we calculate the amplitudes,  $\eta_\sigma$
 and $\eta_{\bar{\sigma}}$, by integrating the square of the eigenfunction
over an integer number of periods.
We test the quality of our fit by integrating the fit equation with the 
numerical eigenfunction over an integer number of periods.
Each value of $a_\sigma$ and $b_\sigma$ depends
on both the total energy value of the state, $\varepsilon$, and the
perpendicular wavevector $k_\perp$.  To obtain a smooth curve for the
conductance, we must have several hundred values of $\varepsilon$.  Depending
upon the critical angles, there will be tens to hundreds of $k_\perp$
corresponding to a single value of $\varepsilon$.
The above procedure must therefore be repeated thousands of times to generate a
single conductance curve.  
Since the $a_\sigma$ and $b_\sigma$ coefficients, and therefore the 
conductance, are functions of these numerically obtained quantities,
they are subject to numerical fluctuations arising from the discretization of
the $k$'s and numerical errors.  

To reduce the influence of numerical fluctuations on the conductance, we use a 
procedure which has both numerical advantages and which better reflects
the physics of the problem.  At any temperature, the current is the 
integral over energies of the conductance:\cite{datta}
\begin{equation}
I(V)=\int G(\varepsilon) (f(\varepsilon-eV)-f(\varepsilon))  d\varepsilon
\label{eq:iv}
\end{equation}
where $G(\varepsilon)=G_\uparrow(\varepsilon)+G_\downarrow(\varepsilon)$ is the 
conductance, $f$ is the Fermi function 
and $V$ is the bias voltage.  
At low $T$, $f(\varepsilon)$ can be approximated %otv3
by a step function, and it is immediately obvious that the elementary
relationship between current and conductance
\begin{equation}
G(V)=\frac{\partial I(V)}{\partial V}
\label{eq:GI}
\end{equation}
is not just a definition but a mathematical identity.

\begin{figure}
%\centering{ \scalebox{1.0}{\includegraphics{fittingprocedure.eps}  } }
\includegraphics [width=3in] {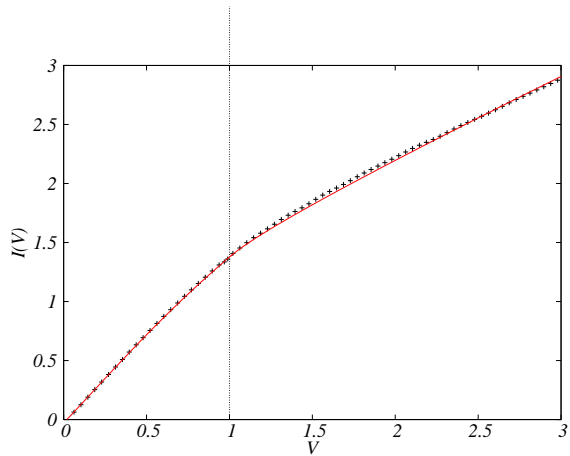}   %otv1
\caption{(Color online) Example of the fitting procedure 
for the normalized current, as described in the text
in connection with \eq(\ref{eq:iv}). The example shown here corresponds to the
case $\Lambda=1$ and $I=0.866$, which will be discused below 
(Figs.~\ref{fig:I866} and \ref{fig:I866gt}). The crosses are the numerical results obtained as
explained in the text and the curves are the fits. The units are
explained in the text} %otv3
\label{fig:smooth}
\end{figure}

It is convenient to numerically evaluate $G(V)$ as follows:  after the set of %otv3
$a$'s and $b$'s have been obtained for all relevant energies and angles, we
apply \eqs(\ref{eq:cond}), (\ref{eq:aa}), and (\ref{eq:iv}) to compute $I(V)$.
%Integrals in \eqs(\ref{eq:aa}) and (\ref{eq:iv}) are computed numerically.
The integral in the first term of 
\eq(\ref{eq:iv}) (the second term is independent of $V$) 
is evaluated numerically by summing all computed %otv3
values of $G(\varepsilon)$ from an energy that is a few $\omega_D$ 
below the the Fermi  energy up  to the bias
voltage, $V$.  The exact value of the lower limit on the integral is not 
important because it will change the resulting value by a constant 
which will disappear 
when \eq(\ref{eq:GI}) is applied.  A non-zero lower limit is advantageous
because the resulting value will depend on a larger number of 
states, thus reducing the effects of numerical fluctuations.
%For $T>0$, care must be taken so that the upper limit does not artificially
%truncate the Fermi functions.
With this procedure, the value of the integral  depends on the
conductance for all voltages below $V$.  This has the effect of reducing
the influence of numerical fluctuations, resulting in  very smooth data,
as illustrated by the example given in Fig.~\ref{fig:smooth}.
The bias %otv3
voltage is given  throughout this work in units of $\Delta_0/e$, so that a 
dimensionless bias voltage of $V=1$ 
corresponds to the gap edge, and  the normalized current
follows  from this
and the dimensionless definition of $G$ discussed below. %otv3
This resulting smooth data
is then fit to an appropriate simple function.  The
fit for the $V>1$ and $V<1$ regions are done separately with the condition  
that the first derivative  of the current be continuous at $V=1$. %otv3
The resulting fit is then numerically differentiated to obtain
the result for $G(V)$.  We follow this procedure for both the forward scattering
and the angularly averaged cases.

\section{Results and Analysis}

%otv paragraph below
In this section, we present our numerical results for the conductance of the
SF bilayers described above, at zero temperature.
The materials that form the junction can be characterized
by four dimensionless  parameters: the exchange 
field parameter $I$, the Fermi wavevector mismatch $\Lambda$, %phbr
the barrier strength $H_B$,
and the dimensionless superconducting coherence length
$\Xi_0 \equiv k_{FS}\xi_0$.  We will vary the $I$ and $\Lambda$ 
parameters over their physically relevant ranges, 
which we take to be $0.1 \le \Lambda \le 2$  and $0\le I \le 1$  %otv3
in our
units. For the coherence length we choose $\Xi_0=50$
and consider both the case of negligible barrier $H_B=0$ and 
that of an intermediate barrier, $H_B=1$.
For stronger interfacial
scattering one of course quickly recovers the well-known and less
interesting standard tunneling results. 
Geometrically, we will use values of the dimensionless
thicknesses $D_F=D_S=1500$. 
%By calculating also conductance curves, at a few values of %phbr
By calculating conductance curves at a few values of
$\Lambda$ and $I$ for smaller and larger lengths, we %phbr
%have found that these values are sufficient to
have found that the lengths used here are sufficient to %phbr
avoid finite size effects: using a smaller sample
size increased numerical fluctuations although
it did not change overall trends and
averages in the
data.  We therefore chose the largest sample size that allowed us to calculate
high quality numerical results in a reasonable amount of computer time.

%otv some changes here too
Following  common convention, dimensionless conductances $G(V)$ are 
normalized to ${e^2}/{h}$.  Thus, for a sample 
with $\Lambda=1$, $I=0$, $H_B=0$ and $\Delta_0=0$, (a homogeneous non-magnetic 
conductor 
in the normal state) we would obtain $G(V)\equiv 1$.  The bias
voltage is in units of $\Delta_0/e$: %otv3 this is repeated should it go out?
thus $V=1$ 
corresponds to the gap edge.
%We have chosen an exchange field $I=0.2=h_0/E_{FM}$ % and 0.866, but wait for it...
%and we use several values of $\Lambda$.
We will present  results for
a range of values of  $\Lambda$, both smaller and larger
than unity, and  consider  values of the exchange field, $I=0.2$, $I=0.5$, and
$I=\sqrt{3}/2\approx0.866$.
Results for $G(V)$ will be given for two cases:  forward scattering 
and angularly averaged scattering.  For the forward scattering case, we 
integrate \eq(\ref{eq:aa}) up to
a small angle of $\pi/30$ for spin-up and spin-down incident particles,
while the angularly averaged results are integrated up to
the maximum possible angle.  
Forward scattering results are most applicable to point contact devices.

\begin{figure}
%\centering{ \scalebox{1.0}{\includegraphics{I0.2.eps}  } }
\includegraphics [width=3.5in] {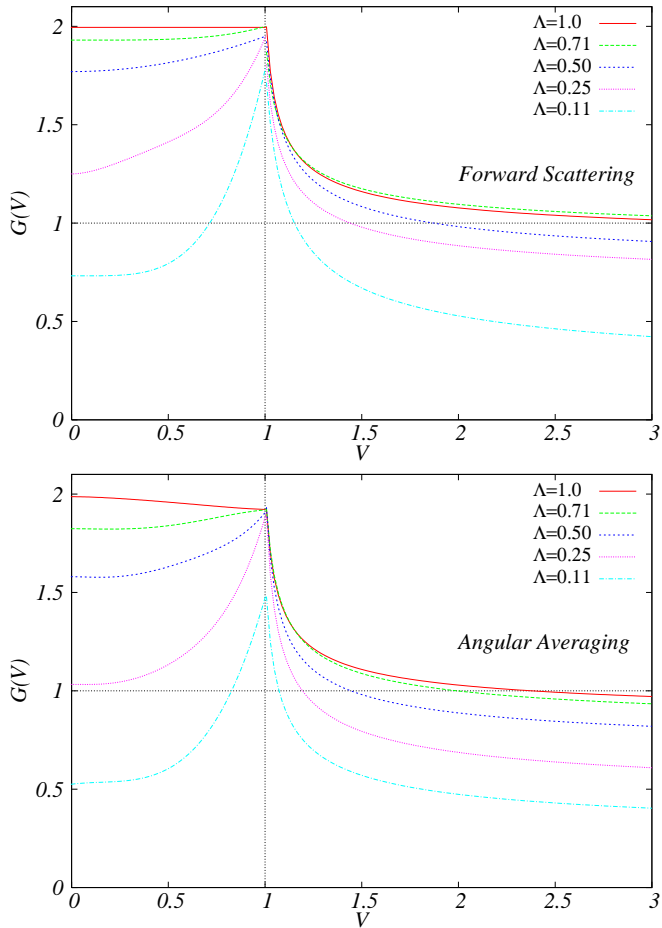}   %phbr
%otv caption rewritten
\caption{(Color online) Results for the dimensionless conductance $G(V)$
(see text) as a function
of bias voltage $V$, given in units of $\Delta_0/e$.
The top panel shows the forward scattering case for $I=0.2$, and  
various $\Lambda\le 1$.  %The conductance, $G(V)$, is in units of $\frac{e^2}{h}$.
%$V$
%is  so that 
%$V=1$ corresponds to the gap edge.
The forward conductance is calculated by averaging over an incident particle 
flux with angles smaller than $\pi/30$.
The bottom panel shows the angularly averaged case at the same values
of $I$ and $\Lambda$.  
The conductances are averaged over angles up to the critical 
angle, $\Omega_\sigma$, for spin-up and spin-down particles.}
\label{fig:I2}
\end{figure}

%otv some revisions here
We consider  first, in Fig.~\ref{fig:I2}, the behavior of $G(V)$ at a fixed
intermediate value of $I$ ($I=0.2$) for several values of $\Lambda$ in the
%range $\Lambda \le 1$ of most experimental relevance. %phbr
range of most experimental relevance, $\Lambda \le 1$. 
%This and subsequent figures  are for $H_B=0$. This is %phbr
This and subsequent figures  are for $H_B=0$,
the experimentally more important case
of a barrier free junction, which we will
study in more detail. Later we will turn to the case where
the tunneling limit is approached by setting $H_B=1$. 
The forward scattering results are shown in the top panel of \fig\ref{fig:I2}. 
%for $I=0.2$ and
%several values of $\Lambda$. 
Of the cases shown in this panel, those in the top
curve ($\Lambda=1$)  serve as a good test
of our technique, since even with a relatively
small exchange field we find results
qualitatively similar to the well established ones obtained by  %otv1
BTK\cite{btk82} at $I=0$.  In the BTK approximation, the conductance at 
zero bias voltage for samples with $I=0$ and $\Lambda=1$ should be equal to $2$ 
due to Andreev reflection, and we see that the
top curve in our plot approaches this limit. 
There, the conductance at $V=0$ is slightly smaller
than $2$, which is easily explained by a slightly suppressed Andreev reflection %phbr
due to the different DOS in the spin-up and spin-down bands at  our
nonzero $I$.
Furthermore, the conductance asymptotically approaches the normal state 
value (unity in these units) at larger bias voltages.   
%The conductance is approximately two for $V<1$, and approaches unity for $V>1$.
We can therefore say that our results approach BTK for $\Lambda=1$
and small $I$.

The  conductance 
is larger 
than unity throughout the sub-gap range 
for all curves in \fig\ref{fig:I2} except  
at the smallest $\Lambda$. In every case, however, we see  that 
the zero bias conductance
is enhanced with respect to the large bias value $G(V \ge 3)$. 
The value of $G(0)$ monotonically decreases with
$\Lambda$, which is reasonable because a smaller $\Lambda$ will lead to stronger
ordinary scattering at the interface and inhibited Andreev reflection.  
If the Andreev reflection and ordinary scattering responded to $\Lambda$ in the
same way, 
%If it were merely a difference in ordinary scattering, 
then the ratio of the zero bias
conductance to the conductance at larger bias, say $V=3$, would be constant.
Instead, we see that the sub-gap enhancement is considerably reduced for 
smaller $\Lambda$, which implies that Andreev reflection is much more sensitive
to $\Lambda$ than ordinary reflection.  
%This is a result of the more complicated scattering
%potential produced by the proximity effect.
The ordering of the conductance curves for $V>1$ is nonmonotonic.  The
order from greatest to least $G$ is $\Lambda=0.71,1.0,0.50,0.25,0.11$.  This is
sensible because the Fermi wavevector for the majority band in the F layer of
the $\Lambda=0.71$ curve is equal to $0.85$ in units of the Fermi wavevector in
S, while that for the $\Lambda=1.0$ curve is $1.2$.  Therefore, the
$\Lambda=0.71$ F layer is slightly better matched to the S than the
$\Lambda=1.0$ layer.

The angularly averaged conductance (bottom panel of \fig\ref{fig:I2}) is 
qualitatively
similar to that found in the forward scattering case.  The values of the 
conductances are
smaller than the forward scattering case, because a smaller
fraction of the incident current (to which everything is normalized) will pass
across the junction.  The zero bias enhancement (with respect
to the large bias limit) is appreciably  less pronounced at %otv1
smaller $\Lambda$.  At $\Lambda=0.11$, the value of $G(0)$ is not much larger
than $G(3)$.  As in the forward scattering case, the values of $G(1)$ 
depend strongly on $\Lambda$.
The $\Lambda=1$ curve in the bottom panel of \fig\ref{fig:I2}
shows a very weak conductance peak at a sub-gap bias.  This is a characteristic
feature\cite{zutic00} of systems for which the junction characteristics are
dominated by the exchange field.
For $\Lambda=0.71$, which introduces a small amount of additional ordinary 
scattering, the conductance peak has
completely disappeared.  
We shall see below that this peak is enhanced with larger $I$.

\begin{figure}
%\centering{ \scalebox{1.0}{\includegraphics{I0.866new.eps}  } }
\includegraphics [width=3.5in] {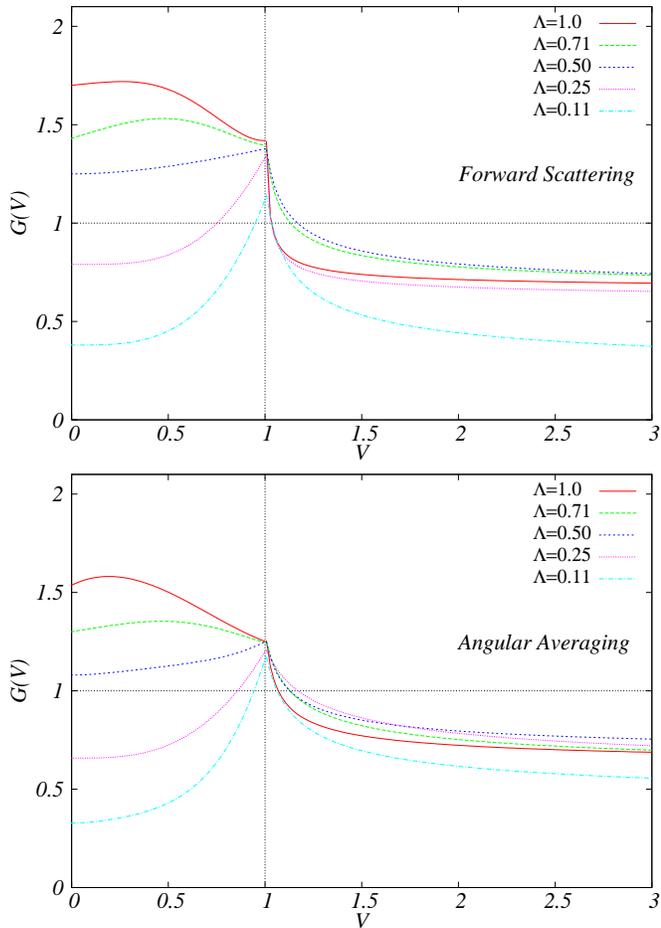}   %phbr
\caption{ (Color online) Results for $G(V)$ at a larger
value of $I$ than in Fig.~\ref{fig:I2}, $I=\sqrt{3}/2\approx0.866$,
and the same range of $\Lambda$.
The top panel shows the forward scattering results and
the bottom panel the angularly averaged values.
The
effects of a strong exchange field are readily apparent when this figure is
compared with \fig\ref{fig:I2} (see text).    
}
\label{fig:I866}
\end{figure}

\begin{figure}
%\centering{ \scalebox{1.0}{\includegraphics{I0.866_Lgt1new.eps}  } }
\includegraphics [width=3.5in] {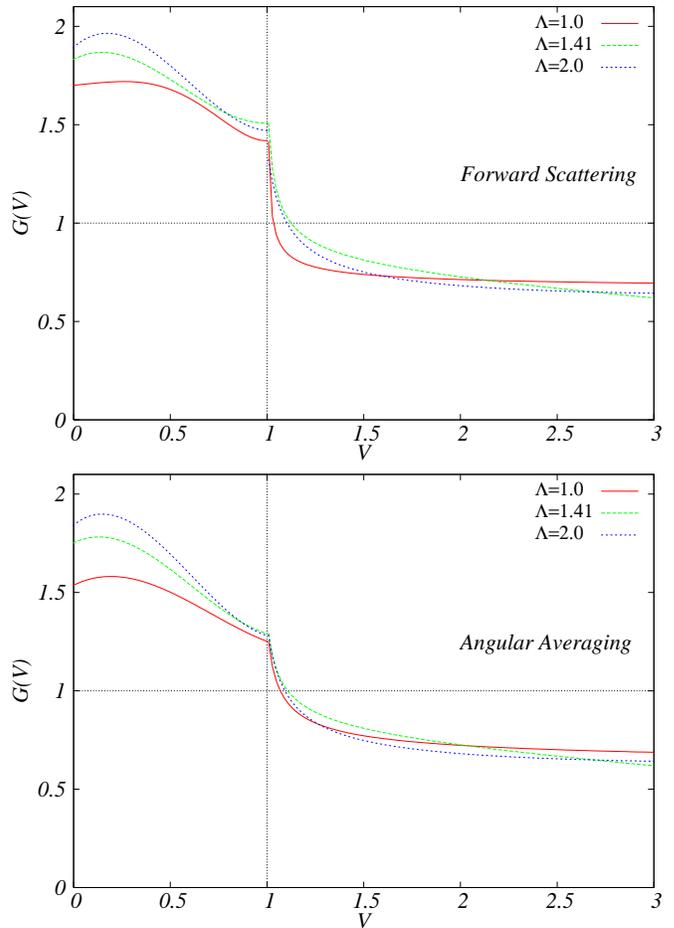}   %phbr
\caption{(Color online) Results for the same case presented in Fig.~\ref{fig:I866}
and with the same panel arrangements,
but for values of $\Lambda$ in the range $\Lambda \ge 1$. See text for discussion.} %otv1
\label{fig:I866gt}
\end{figure}

The results for a very strong exchange field ($I=0.866$) are shown in
Figs.~\ref{fig:I866} and
\ref{fig:I866gt}. The first of these figures covers
the case where $\Lambda <1$. The strong exchange field
makes many features that are already present in \fig\ref{fig:I2} %phbr
more dramatic.
When present, the peaks at sub-gap biases are proportionally
much higher.  In the case of $\Lambda=0.25$, there is now 
%no sub-gap enhancement of the conductance, and for $\Lambda=0.11$ the %phbr
a very weak zero bias enhancement in the forward scattering case, and no %otv1 
such %otv1 
enhancement in the angular averaging case.  %phbr
For $\Lambda=0.11$, the  %phbr
suppression of the conductance for sub-gap bias voltages is very strong.
As in the $I=0.2$ case, the ordering of the conductance curves for $V>1$ is 
nonmonotonic: the order from greatest to least is 
$\Lambda=0.50,0.71,1.0,0.25,0.11$.  As before, the ordering is dictated by the 
degree to which the Fermi wavevector of the majority band in F matches the 
Fermi wavevector in S.  In all cases, the value is small because the minority 
band contributes a minuscule amount to the conductivity. 

%otv1 edited below (was changed but not marked with %phb)
The next figure (Fig.~\ref{fig:I866gt}) is for $I=0.866$, as in 
Fig.~\ref{fig:I866}, but with $\Lambda \ge 1$.  Such large values of 
$\Lambda$ might be experimentally 
the case only if the superconducting material were some oxide material
with s-wave pairing, but it is nevertheless of theoretical interest.
For the two values of $\Lambda >1$ considered there we see how the 
previously observed  trends 
in $\Lambda$ continue when the mismatch is in the opposite direction. 
We see that the subgap behavior is monotonic with a marked peak at nonzero bias,
and that the conductance is nonmonotonic in $\Lambda$ for $V>1$.

\begin{figure}
%\centering{ \scalebox{1.0}{\includegraphics{fig6.eps}  } }
\includegraphics [width=3.5in] {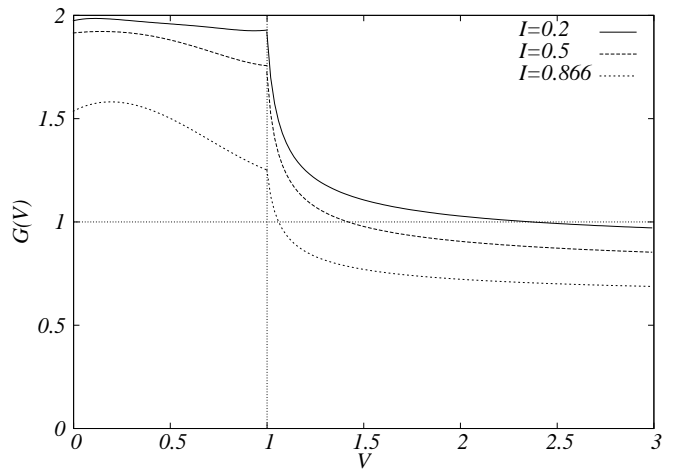}   %phbr
\caption{Angularly averaged values of $G(V)$ for various $I$, 
at the same $\Lambda=1.0$. 
%The units of the conductance and bias voltages are a given in previous figures.
This figure shows (see text) %otv1
that stronger $I$ suppresses both ordinary
scattering and %otv1
Andreev reflection.  Note that $G(0)/G(3)\approx2$
for all values of I shown here, and that %otv1
 $G(0)/G(1)$ increases with increasing $I$.}
\label{fig:L1ang}
\end{figure}

In \fig\ref{fig:L1ang}, we consider the effect of
changing $I$ at constant $\Lambda$.
The mismatch is held to  $\Lambda=1$ and the
exchange field is taken from a moderate $I=0.2$ to a strong
$I=0.866$.
%(so that $I^2=0.75$).  
We chose to show
the angularly averaged case because the sub-gap conductance peak is more
apparent than in the forward scattering case.  The $I=0.2$ curve, 
which was discussed above, % and is presented here only for comparison; it 
appears nearly flat when
in the company of curves with larger $I$.  For $I=0.5$, the
conductance peak is a little more pronounced. %otv can hardly see it?
Overall, the conductance is
smaller as $I$ increases because a larger exchange field leads to more 
poorly matched Fermi wavevectors from F to S.
The $I=0.866$ conductance curve is the most dramatic.  
The value at zero bias is proportionally much larger than $G(1)$ than for $I=0.5$ or $I=0.2$.  
%The conductance within the gap is suppressed to a much greater degree than 
%the conductance outside of the gap.  
The ratio of the zero bias conductance to the conductance at larger bias
voltages is approximately two for all values of $I$ shown here.  This
contradicts the common assertion that the reduction in the minority band
DOS invariably leads to a smaller Andreev reflection.

\begin{figure}
%\centering{ \scalebox{1.0}{\includegraphics{fig7.eps}  } }
%\centering{ \scalebox{1.0} {\includegraphics{newgv1.ps} } }
\includegraphics[width=3.5in] {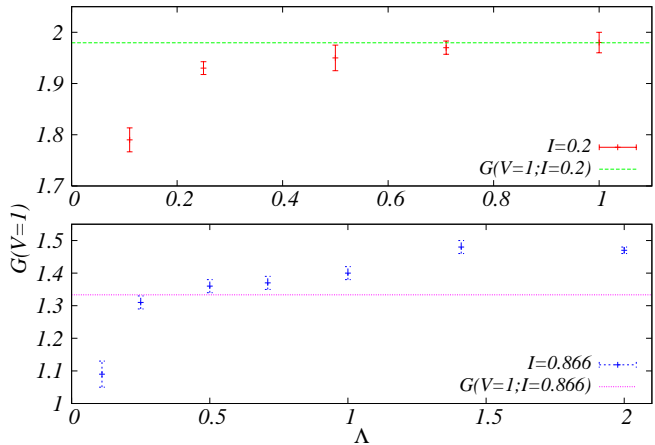}  %phbr
\caption{(Color online) 
%The top panel 
Plots of $G(I;V=1)$, the value of the dimensionless conductance at $V=1$
in
the forward scattering case
for $I=0.2$ (top panel) %, top horizontal scale) %phb3
 and  $I=0.866$ (bottom panel).  %, bottom horizontal scale). %phb3
The horizontal lines
labeled by $G(V,I)$  show the result from the non-self-consistent formula,
\eq(\ref{eq:GV1}). The points represent the numerical
results from the present self-consistent study, at the values of $I$
indicated in the legend.  
The error bars represent the 
uncertainty in the numerically obtained values (see text). 
Note the different horizontal range included in the upper and
lower panels. 
%The lower panel shows $G(I;V=1)$ as a function of $I$.  The
%dashed curve is the non-self-consistent result.
}
\label{fig:GV1}
\end{figure}

In all figures shown thus far, the  cusp in the conductance for bias voltages 
corresponding to the gap edge is dependent on both $\Lambda$ and $I$.  
The non-self-consistent analysis\cite{zutic00} 
predicts a value of $G(1)$ that is dependent on $I$ but independent of
$\Lambda$, namely: 
\begin{equation}
G(V=1;I)=\frac{4(1-I^2)^{1/2}}
{1+(1-I^2)^{1/2}}.  
\label{eq:GV1}
\end{equation}
%We find in the self-consistent analysis that for  $\Lambda=1$, $G(1)\approx2$, 
%while for $\Lambda=0.11$, $G(1)\approx 1.5$, with the others ordered
%monotonically in between.  
The non-self-consistent result
depends very strongly on the
assumption that the order parameter is independent of $\Lambda$,
an assumption which is not  valid, particularly  near the interface,
when the order parameter is calculated 
self-consistently.\cite{hv02a} 
%Further discussion will be presented in the context of \fig\ref{fig:GV1}.
Therefore one should look for violations of this relation in the correct,
self-consistent calculation. 
Figure~\ref{fig:GV1} is a direct comparison between the non-self-consistent
formula\cite{zutic00} for $G(V=1;I)$, \eq(\ref{eq:GV1}), and the self-consistent
results.  
%The top panel 
The figure shows $G(V=1;I)$ for $I=0.2$ (top panel) 
and $I=0.866$ (bottom panel). The results for $I=0.2$
are plotted in the range of $\Lambda$ included in Fig.~\ref{fig:I2} and
those for $I=0.866$ in the more extended range included
in Figs.~\ref{fig:I866} and \ref{fig:I866gt}.
The
non-self-consistent results, which are, as explained
above, indeppendent of $\Lambda$, are shown as 
the horizontal lines.  The self-consistent %phbr
results are the data points.  The error bars represent the numerical
uncertainty which arises because $G(V)$ has a sharp cusp at  $V=1$.
This discontinuous first derivative at $V=1$ makes this the most
difficult value of the conductance to obtain numerically.  
Even with the numerical error, the
difference  between the two methods is apparent:  there is a
clear  increasing trend 
in the self-consistent results as a function of $\Lambda$ in contrast
with the $\Lambda$-independent value
predicted by the non-self-consistent result.  
This discrepancy is stronger at larger $I$ 
and it affects (see bottom panel) both the small and large values
of $\Lambda$, particualrly the latter ones. 
This large discrepancy here is due, as mentioned
above, to the fact that the non-self-consistent result
relies on the assumption that $\Delta$ is independent of $\Lambda$.

In general, we find that for all values of $I$ the exact results are in
fair agreement with those obtained within the non-self-consistent
approximation near $\Lambda=1$ but that very considerable deviations occur
in all cases for values of this parameter both  smaller and  larger than
unity. Although in some intuitive way this may seem to make obvious
%sense, it is actually nontrivial, since after all the self-consistent result
sense, it is actually nontrivial, since the self-consistent result %phbr
for $\Delta(z)$ near the interface is never equal to the non-self
consistent (that is, the bulk) value. It seems, though, that the effect
on this on the result for $G(V=1)$ is minimized when the mismatch is 
at a minimum. This is, however, rarely the experimental situation: for
most materials of experimental interest there is considerable mismatch.

In the non-self-consistent treatment, it was found that the Fermi wavevector
mismatch, $\Lambda$, has an important effect on the conductance
curves.\cite{zutic00} In the results found  in this section, we have shown 
that the  effect of
$\Lambda$ is even greater when self-consistently is included.
It is a not unusual experimental practice\cite{leighton04,leighton06} to
characterize conductance curves by using only a
polarization parameter $P$, an interfacial barrier strength $H_B$, %$Z_B$, %phbr
and $\Delta_0$ as fit parameters.
In light of the results  in Ref.~\onlinecite{zutic00}, confirmed and reinforced here,
neglecting $\Lambda$ is a deplorable practice, and can 
easily lead to spurious results.
A much sounder procedure %otv1
would be to fit results for different samples using  
$\Lambda$, $H_B$ and $I$ as  fitting %phbr changed Z_B to H_B
parameters, taking $\Delta_0$ as the value of the bulk material and
disentangling the effects of  $\Lambda$  and $H_B$ by remembering that %phbr changed Z_B to H_B
the latter, but not the former, will change from junction to junction.  %otv1 !!
%using some
%other method to characterize $Z_B$.

\begin{figure}
%\centering{ \scalebox{1.0}{\includegraphics{tunneling.ps}  } }
\includegraphics[width=3.5in] {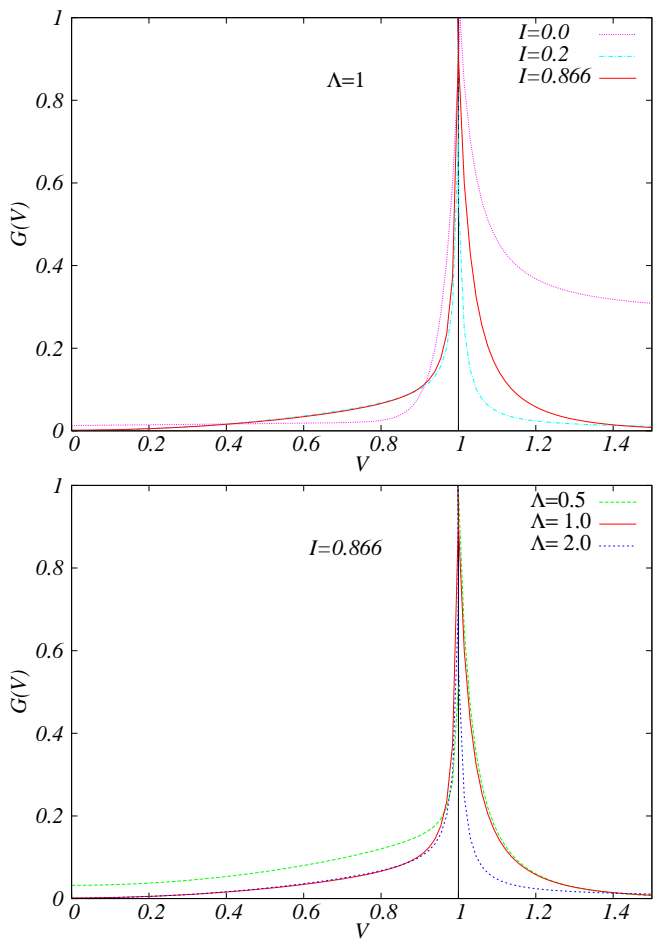}   %phbr
\caption{ (Color online)
The conductance at interface scattering value $H_B=1$,
near  the tunneling regime. All results 
correspond to the forward scattering case.
The top panel 
shows $G(V)$ for three values of $I$ at $\Lambda=1$,
while the bottom panel shows $G(V)$ at $I=0.866$
for three values of $\Lambda$.  
}
\label{fig8}
\end{figure}

All of  the previous results shown have been obtained in the regime
of most experimental interest where the interfacial scattering parameter
$H_B$ is negligible. In the next figure, Fig.~\ref{fig8}, we present results
for $H_B=1$, when the tunneling regime $(H_B \gg 1)$ is approached, as can
be seen from the very different shape of the curves. The two panels in the
figure display both the dependence of the results on $I$ at constant
$\Lambda$ (top panel) and the dependence on $\Lambda$ at constant $I$ (bottom
panel). In general, these dependences are not strong:  this is
because in the true
tunneling regime $G(V)$ simply reflects the superconducting
DOS and, as one approaches
this  regime,  this should of course result in a weaker
$\Lambda$ dependence.  This is well known to 
experimentalists: to obtain reliable measurements of 
quantities such as the polarization parameter it is always preferable to work with samples
that have small interfacial scattering. In other words, a larger $H_B$ 
inhibits Andreev reflection and the proximity effects.

Nevertheless, at $H_B=1$ some definite trends can be ascertained and
definite statements can be made. %phb4 
In the top panel of Fig.~\ref{fig8} one can see
that the $I$ dependence at constant $\Lambda$ is still relatively strong,
particularly in the region $V>1$. It is remarkable that the behavior
with $I$ is strongly nonmonotonic  at larger voltages and also, although
much more weakly, at $V<1$. This makes interpolation schemes very doubtful %phbr
in interpreting experimental data. In the bottom panel, corresponding
to a strong magnet, we see that the mismatch dependence is weak 
in the region $V>1$ for $\Lambda \le 1$ but quite noticeable  for $\Lambda >1$.
In the region $V<1$ the situation is exactly the opposite: the curves %phb4
corresponding to $\Lambda=1$ and $\Lambda=2$ nearly coincide, while that
for smaller $\Lambda$ is clearly different. This nonmonotonic behavior
contrasts again with that found in non-self-consistent results (see for
example panel (b) of Fig.~2 in Ref.~\onlinecite{zutic00}) which, for
similar values of $I$, vary monotonically\cite{pardef} with mismatch
in the same way over the whole range of $V$.

\section{Conclusions}

We have introduced a method for calculating conductances in an F/S
bilayer within a fully
self-consistent microscopic model.  %phbr next few sentences moved from later
Many of the features that we find are seen in
experimental work:  there is a sub-gap enhancement of the %otv3
conductance, the conductance does approach the normal state value for larger
bias voltages, and there is a cusp at bias voltages of unity.
Most important, we show that, as already indicated
by the non-self-consistent results, detailed experimental analysis
(in particular the extraction of the spin polarization) is impossible
if one does not take into account separately the effects of mismatch
and those of  barrier scattering.

The features of the conductance curves
agree only qualitatively with  those obtained via non-self-consistent
procedures. We find that there
are strong
quantitative differences.  
One of the most obvious is that we find a strong dependence
% phb4 changed $G(V=\Delta)$ to $G(V\equiv\Delta_0)$
of $G(V\equiv\Delta_0)$ on the parameter $\Lambda$, %phbr
which characterizes the Fermi wavevector mismatch. In the
non-self-consistent approach, $G(V\equiv\Delta_0)$ is analytically
found\cite{zutic00} to be independent of $\Lambda$. 
In both treatments, the dependence of $G(V\equiv\Delta_0)$ on $I$ is similar in trend
but quite different in detail. %otv1 %phbr
%The dependence of  $G(1)$ on $I$ is similar in trend, but it differs
%in detail. 
The conductance is
reduced for larger $\Lambda$ and larger $I$.  We find a 
sub-gap conductance peak %phb3 (note fig. 4:  peak is @ finite bias)
%zero bias conductance
for strong $I$ and $\Lambda$ close to unity. %otv3
All of this 
indicates that, while the non-self-consistent %phbr
approach is a good tool to help us understand qualitatively
some of the features of SF transport, 
a fully self-consistent
approach is needed to properly model experimental data.

This paper represents
merely the first step in studying SF bilayers using a fully 
self-consistent pair amplitude.  
Future work may involve the addition of normal metal electrodes at the
boundaries of the sample, allowing us to explore the effects of finite F and S
widths.  Detailed studies of the relationship between the local densities of
states (DOS) and the conductance are also desirable.  Finally, 
the effects of finite temperature
should be explored.  

\acknowledgments
 
We are very grateful to Klaus Halterman for many discussions on the technical
aspects of this numerical work and to Igor \u{Z}uti\'{c} for numerous
conversations on this problem.


\begin{thebibliography}{0}
\bibitem{pardef} The mismatch parameter $L_0$ used in Ref.~\onlinecite{zutic00}
is related to $\Lambda$ as $L_0^2=1/\Lambda$.
\end{thebibliography}
\end{document}